\documentclass[pdflatex,sn-mathphys-num]{sn-jnl}

\usepackage[T1]{fontenc}

\usepackage{graphicx}%
\usepackage{multirow}%
\usepackage{amsmath,amssymb,amsfonts}%
\usepackage{amsthm}%
\usepackage{mathrsfs}%
\usepackage[title]{appendix}%
\usepackage{xcolor}%
\usepackage{textcomp}%
\usepackage{manyfoot}%
\usepackage{booktabs}%
\usepackage{algorithm}%
\usepackage{algorithmicx}%
\usepackage{algpseudocode}%
\usepackage{listings}%
\usepackage{comment}
\usepackage{braket}
\usepackage{qcircuit}
\usepackage{physics}

\theoremstyle{thmstyleone}%
\theoremstyle{thmstyletwo}%

\theoremstyle{thmstylethree}%

\begin{document}

\title[]{Selective Feature Re-Encoded Quantum Convolutional Neural Network with Joint Optimization for Image Classification}

\author[1]{\fnm{Shaswata Mahernob} \sur{Sarkar}}\email{smsarkar@ieee.org}

\author[1]{\fnm{Sheikh Iftekhar} \sur{Ahmed}}\email{sheikh.iftekhar@ieee.org}

\author[2]{\fnm{Jishnu} \sur{Mahmud}}\email{jishnu.mahmud@ieee.org}

\author*[1]{\fnm{Shaikh Anowarul} \sur{Fattah}}\email{fattah@eee.buet.ac.bd}

\author[3]{\fnm{Gaurav} \sur{Sharma}}\email{gaurav.sharma@rochester.edu}

\affil[1]{\orgdiv{Department of Electrical and Electronic Engineering}, \orgname{Bangladesh University of Engineering and Technology}, \orgaddress{\city{Dhaka}, \postcode{1000}, \country{Bangladesh}}}

\affil[2]{\orgdiv{Department of Computer Science and Engineering}, \orgname{Brac University}, \orgaddress{\city{Dhaka}, \postcode{1212}, \country{Bangladesh}}}

\affil[3]{\orgdiv{Department of Electrical and Computer Engineering}, \orgname{University of Rochester}, \orgaddress{\city{Rochester}, \country{USA}}}

\abstract{
\renewcommand{\thefootnote}{}%
\footnotetext{\textit{This is a pre-print of a manuscript that has been submitted for peer review.}}%
\renewcommand{\thefootnote}{\arabic{footnote}}%
Quantum Machine Learning (QML) has seen significant advancements, driven by recent improvements in Noisy Intermediate-Scale Quantum (NISQ) devices. Leveraging quantum principles such as entanglement and superposition, quantum convolutional neural networks (QCNNs) have demonstrated promising results in classifying both quantum and classical data. This study examines QCNNs in the context of image classification and proposes a novel strategy to enhance feature processing and a QCNN architecture for improved classification accuracy. First, a selective feature re-encoding strategy is proposed, which directs the quantum circuits to prioritize the most informative features, thereby effectively navigating the crucial regions of the Hilbert space to find the optimal solution space. Secondly, a novel parallel-mode QCNN architecture is designed to simultaneously incorporate features extracted by two classical methods, Principal Component Analysis (PCA) and Autoencoders, within a unified training scheme. The joint optimization involved in the training process allows the QCNN to benefit from complementary feature representations, enabling better mutual readjustment of model parameters. To assess these methodologies, comprehensive experiments have been performed using the widely used MNIST and Fashion MNIST datasets for binary classification tasks. Experimental findings reveal that the selective feature re-encoding method significantly improves the quantum circuit's feature processing capability and performance. Furthermore, the jointly optimized parallel QCNN architecture consistently outperforms the individual QCNN models and the traditional ensemble approach involving independent learning followed by decision fusion, confirming its superior accuracy and generalization capabilities.}


\maketitle

\section{Introduction}\label{intro}
Machine learning (ML) and deep neural networks have become pivotal in modern science and technology, excelling at producing precise predictions from a large amount of data. Convolutional Neural Networks (CNNs) are designed to process gridded data such as images \cite{krizhevsky2012imagenet}. CNNs utilize kernels to learn spatial feature hierarchies, especially suitable for image classification tasks. CNNs are widely used in medical diagnosis \cite{hu2020learning, alqahtani2024cnx}, face recognition \cite{zeng2021survey, song2020similar}, autonomous driving \cite{rawashdeh2021drivable}, security surveillance \cite{dandamudi2020cnn}, and so on. However, classical architectures such as CNNs face limitations when managing highly complex data and computationally intensive tasks, as computational demands increase exponentially with the dataset size and problem complexity. Furthermore, as semiconductor fabrication approaches its limits in the post-Moore law era, concerns grow about the sustainability of traditional computational methods \cite{urhobo2024comes}. 

These challenges open the door to quantum computing (QC) and quantum machine learning (QML), new paradigms for tackling computational problems intractable for classical systems \cite{fedorov2022quantum, brassard1998quantum}. Several recent QML algorithms have been shown to theoretically outperform their best known classical counterparts \cite{9274431, biamonte2017quantum, brandao2019quantum, schuld2016prediction}. By utilizing quantum properties such as entanglement and superposition, QML enhances classical ML components such as perceptrons and convolutions. Altaisky et al. \cite{altaisky2001quantum} pioneered this by introducing the quantum perceptron, replacing the classical DNN layers with qubits and unitary operators. This idea was further developed by \cite{schuld2015simulating} and \cite{cao2017quantum}, showing that quantum perceptrons are capable of handling multiple inputs simultaneously in superposition, although they require twice the number of qubits compared to the size of input data. A quantum feed-forward network is proposed by Wan et al. \cite{wan2017quantum} that uses ancillary qubits and gradient descent for training without quantum state measurement. Later, Zhao et al. \cite{zhao2021qdnn} introduced the Quantum Deep Neural Network (QDNN), analogous to a classical DNN, consisting of a quantum input layer, hidden layers, and an output layer.

The current phase of quantum computing, called the Noisy Intermediate Scale Quantum (NISQ) era, struggles with issues such as qubit crosstalk, quantum decoherence, and imperfect gate calibration, which limit scalability \cite{patterson2019calibration}. However, breakthroughs such as the achievement of 99\% fidelity in silicon-based quantum computing \cite{mkadzik2022precision} indicate promising progress toward nearly error-free quantum computing. QML merges quantum algorithms with classical ML to increase model efficiency and capability. One such powerful framework is the Parameterized Quantum Circuit (PQC), which consists of adjustable gate parameters, where classical information can be encoded into quantum states and represented in Hilbert space \cite{ cerezo2021variational, benedetti2019parameterized, 9144562}. These circuits process quantum states, compute a post-measurement cost function, and optimize parameters classically through gradient descent \cite{li2017hybrid}. However, encoding high-dimensional data requires a significant quantity of qubits. The limited qubit availability in current hardware necessitates dimensionality reduction techniques, such as PCA and autoencoders, before starting quantum processing. Another challenge in PQCs is the barren plateau phenomenon, which causes a vanishing gradient problem. It worsens with the increasing number of qubits and disrupts the training process  \cite{mcclean2018barren}. Hierarchical structures such as quantum convolutional neural networks (QCNNs) can significantly mitigate this issue by exponentially reducing qubits with circuit depth \cite{grant2018hierarchical}. Pesah et al. \cite{pesah2021absence} demonstrated QCNNs’ ability to alleviate barren plateaus through effective gradient scaling. The architectures proposed by \cite{cong2019quantum} consist solely of quantum convolutional and quantum pooling layers, analogous to classical CNNs. Later, Hur et al. \cite{hur2022quantum} explored fully parameterized QCNN models for binary classification, testing various ansatzes and encoding strategies.

Some studies proposed hybrid designs that integrate both quantum and classical layers. Henderson et al. \cite{henderson2020quanvolutional} introduced the quanvolutional layer, which produces feature maps similar to a classical convolutional layer by using random quantum circuits. This quantum layer processes spatially-local subsections of images, with the quantum circuit's weights remaining untrained. The effects of different sliding modes of quantum filters and parameter sharing were explored in the study by \cite{shi2024quantum}, finding that complex filter orientations and unshared parameters can improve performance. In a study by Easom et al. \cite{easom2022efficient}, a deep quantum neural network utilizing a single qubit was introduced for image classification that follows conventional CNN methods, leading to a decrease in the parameter count. A classical to quantum transfer learning framework was implemented by Kim et al. \cite{kim2023classical}, where a classical CNN was trained on the Fashion MNIST dataset at first, and then the pre-trained CNN was used as a feature extractor for the MNIST dataset. Mahmud et al. \cite{mahmud2024quantum} proposed novel quantum interaction layers and used ancilla qubits for measurement, setting a new benchmark for binary classification on MNIST and Fashion MNIST datasets. However, employing ancilla qubits unnecessarily increases the overall qubit count computational overhead.  

These conventional QCNN models encode classical data into quantum states only once at the initialization of the circuit. However, re-encoding these classical data in the QCNN during quantum feature processing remained largely unexplored, which might refine the internal computations to focus on the intended target state. In this work, a selective feature re-encoding technique has been implemented for the QCNN framework, which effectively selects the most significant features and encodes them in the circuit's intermediate phases. This approach guides the QCNN toward a more appropriate and optimal solution in the Hilbert space. Additionally, existing work on QCNN often relies on features derived from a single classical feature extraction technique. However, the dependency on a single type of feature might not generalize well when the dataset is varied. In this study, two popular and effective feature extraction techniques, PCA and autoencoders, are utilized in separate QCNNs within a unified training pipeline. A novel interaction block is proposed, which interconnects the quantum feature spaces from individual models and allows the interchange of information from distinct features. In classical networks, ensemble learning is a popular approach for this purpose, where each model is individually trained and their outputs are fused for the final result. In contrast, the proposed approach enables quantum domain interaction between the quantum states of each model, leading to a joint optimization of the entire network, achieving richer feature representation. The outputs from individual models are combined after the quantum interaction. The usefulness of this method has been validated by comparing its performance with the individual training scheme and the conventional ensemble strategy. The major contributions of this study are as follows:\\

\begin{enumerate}
    \item Selective feature re-encoding layers are proposed within the QCNN architecture that enable the model to reach a more optimal solution and improve classification performance.
    
    \item A parallel mode integration of two QCNN models is proposed, with one QCNN model using PCA-extracted features and the other utilizing autoencoder features.
    
    \item Two strategies are employed to combine the QCNN models: a joint optimization approach and an ensemble learning approach. In the joint optimization, the QCNN models participate in a combined training pipeline using a shared loss function. In the ensemble learning approach, the output measurements of each independently trained QCNN model are merged to produce final predictions.

    \item Extensive experiments are carried out to comprehensively evaluate the proposed approaches by using multiple binary combinations from the two most widely used MNIST and Fashion MNIST datasets. The results indicate that the proposed feature re-encoding strategy improves the performance over conventional QCNN. Also, the joint optimization approach outperforms individual QCNNs and the ensemble approach.  
\end{enumerate}

\section{QCNN with Proposed Selective Feature Re-Encoding Strategy} \label{sec3}

\begin{figure}[h!]
    \centering
    \includegraphics[width=0.7\linewidth]{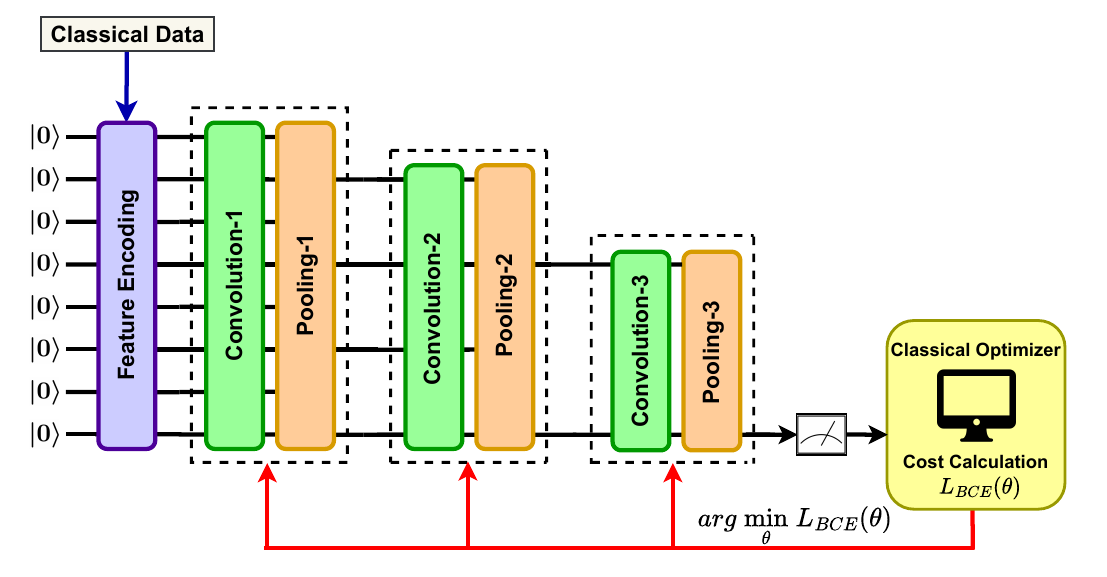}
    \caption{A structural diagram of conventional QCNN with 8 input qubits. It consists of an initial data encoding layer followed by three conv-pool operation stages, resulting in a single qubit. The measurement outcome from the last qubit 
    is used for cost calculation. Finally, a classical optimizer updates the trainable weights of the circuit using the gradient descent technique.}
    \label{fig:conventional_qcnn}
\end{figure}

\begin{figure}[h!]
    \centering  \includegraphics[width=1\linewidth]{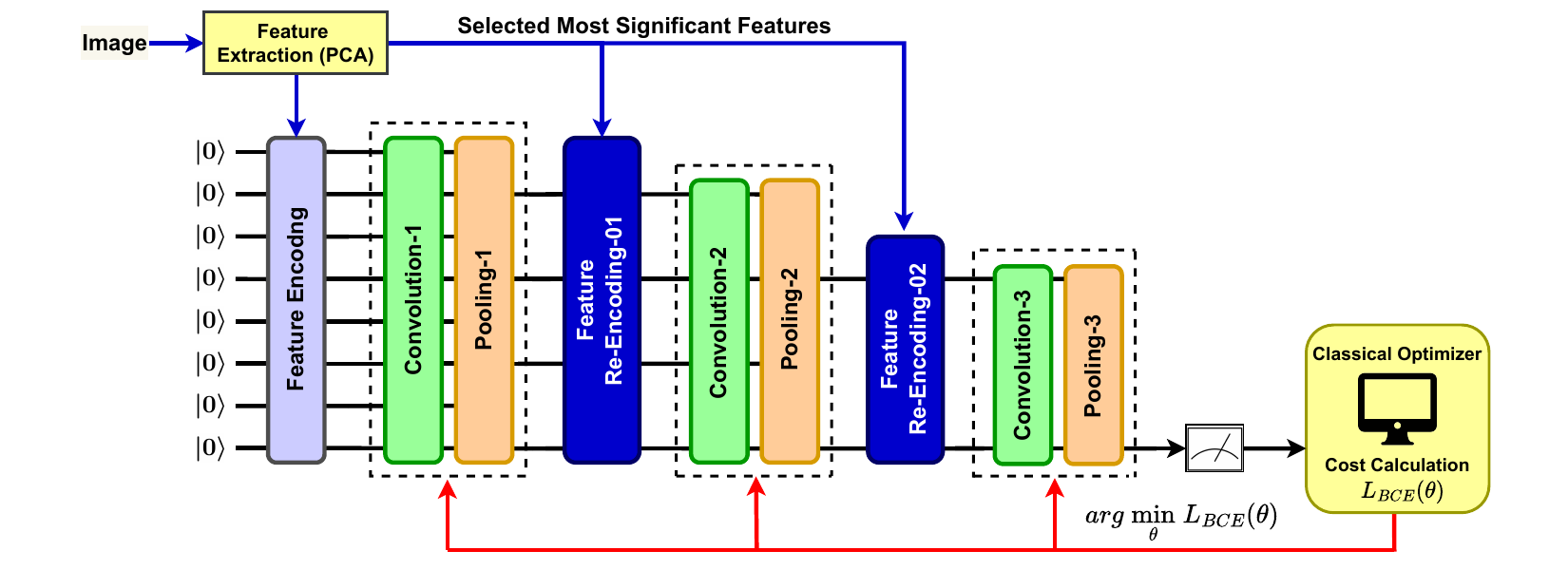}
    \caption{Proposed QCNN architecture with selective feature re-encoding strategy for an 8-qubit system. In addition to having the initial feature encoding layer, there are two additional re-encoding layers (blue blocks), one each after the pooling layer-1 and the pooling layer-2. The most significant selected features extracted through PCA are fed into the quantum circuit through these re-encoding layers. The third pooling operation results in the final single qubit, which is only used for measurement.}
    \label{fig:reencoded_qcnn}
\end{figure}

Quantum Convolutional Neural Networks (QCNNs) are specialized variational quantum circuits, where parameterized quantum convolutional and pooling ansatzes are employed in a systematic design for effective processing of quantum features for a target task. The architecture starts with the encoding of classical data into the quantum Hilbert space, followed by alternating convolution and pooling layers, and concludes with a measurement layer. The convolutional layer preserves translational invariance in the data, while the pooling layer extracts a fraction of operational qubits, reducing circuit dimension. A general schematic of architecture is illustrated in Fig. \ref{fig:conventional_qcnn}. In conventional QCNNs, the classical to quantum feature encoding operation is performed only once at the start, and no further manipulation is done within the quantum circuit from the classical domain. These initially encoded quantum features are processed through unitary operations having randomly initialized gate parameters, where the circuit may lose the distinguishing characteristics of the input data for classification with increasing depth. Specifically for QCNNs, the quantum feature space dimensionality goes through a direct reduction after pooling layers, losing crucial information content. This can hinder the effective exploration of the optimal solution space within the Hilbert space, thereby affecting classification performance. To compensate for this and fine-tune the target focus of quantum processing within the circuit, the most significant input features can be re-encoded into the Hilbert space after each pool operation. Here, the same initial features cannot be reused in QCNN since the number of operational qubits for data re-encoding is reduced through pooling. Hence, a feature extraction or transformation strategy is required that allows for picking fewer yet significant features to be re-encoded into the reduced set of qubits, thus contributing as an effective guide towards the optimal solution. In this study, a new selective feature re-encoding method is proposed for the QCNN architecture to achieve these goals.

\subsection{QCNN Architecture with Feature Re-Encoding}\label{feature_reencod}

Following the traditional configuration, the initial encoding layer encodes classical features as quantum features into the circuit, which are subjected to the first conv-pool operation. The pooling layer reduces operational qubits, causing a shrinkage in the quantum feature space and potential information loss. At this stage, the architecture of Fig. \ref{fig:conventional_qcnn} is modified by applying a feature re-encoding layer to the reduced qubits after pooling. The most significant selected features are reintroduced into the circuit to compensate for information loss and as guidance towards the solution space. This strategy is repeated after each pooling layer with the same purpose, except for the last conv-pool stage, after which a single qubit remains for measurement only, and no re-encoding is applied to it. The improved QCNN architecture with feature re-encoding strategy for an 8-qubit system is illustrated in Fig. \ref{fig:reencoded_qcnn}. This architecture differs from the conventional QCNN by introducing extra feature re-encoding layers after each of the intermediate pooling layers. The effective qubit number gets halved to 4 after the first pooling; therefore, the first feature re-encoding layer is applied to retain the target direction of quantum operations. Similarly, the second pooling returns 2 operational qubits, and the second feature re-encoding layer is introduced with the same purpose. After the third conv-pool stage, only the $8^{\textrm{th}}$ qubit remains operational, which is subjected to a measurement operation for predicting the outcome.

\begin{figure}[h!]
    \centering
    \resizebox{0.45\linewidth}{!}{
    \centering
    \Qcircuit @C=1em @R=1em {
    \lstick{} & \gate{R_y(\theta_1)} & \ctrl{1} & \gate{R_y(\theta_3)} & \ctrl{1} & \gate{R_y(\theta_5)} & \qw \\
    \lstick{} & \gate{R_y(\theta_2)} & \targ    & \gate{R_y(\theta_4)} & \targ    & \gate{R_y(\theta_6)} & \qw
    }
    }
    \caption{Convolutional Ansatz-1 [Special Orthogonal $SO(4)$] containing 6 trainable parameters ($\theta_1 \to \theta_6$). $R_y(\theta_i)$ denotes rotation around $y$-axis of the Bloch sphere by angle $\theta_i$, and CNOT gates change entanglement between the qubits.}
    \label{fig:SO4_ansatz}
\end{figure}
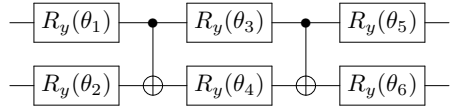

In this study, two different convolutional ansatzes are explored to implement the convolutional layers. The first one is shown in Fig. \ref{fig:SO4_ansatz}, denoted as Convolutional Ansatz-1. It consists of a sequence of parameterized single-qubit rotation gates $R_y(\theta_i)$, and entangling operations using CNOT gates. This 2-qubit quantum circuit is capable of implementing any arbitrary $SO$(4) gate \cite{wei2012decomposition}. 

\begin{figure}[h!]
    \centering
    \resizebox{0.75\linewidth}{!}{ 
    \Qcircuit @C=0.9em @R=1.3em { 
        & \gate{U_3(\theta_1, \theta_2, \theta_3)} & \targ & \gate{R_z(\theta_7)} & \ctrl{1} & \qw & \targ & \gate{U_3(\theta_{10}, \theta_{11}, \theta_{12})} & \qw \\
        & \gate{U_3(\theta_4, \theta_5, \theta_6)} & \ctrl{-1} & \gate{R_y(\theta_8)} & \targ & \gate{R_y(\theta_9)} & \ctrl{-1} & \gate{U_3(\theta_{13}, \theta_{14}, \theta_{15})} & \qw \\
    }
    }
    \caption{Convolutional Ansatz-2 [Special Unitary $SU(4)$] comprising 15 trainable parameters ($\theta_1 \to \theta_{15}$). Here, $R_j(\theta_i)$ denotes rotation around the $j$-th axis of the Bloch sphere by angle $\theta_i$, and $U_3$ gates can perform any arbitrary single-qubit rotation around the Bloch sphere.}
    \label{fig:SU(4)}
\end{figure}
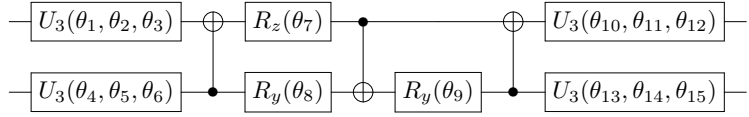

The Convolutional Ansatz-2 is illustrated in Fig. \ref{fig:SU(4)}, comprising 3 CNOT gates and 15 elementary single-qubit gates from the family $\{R_{x}, R_{y}, R_{z}\}$. This circuit can achieve a general two-qubit quantum computation up to a global phase, thus representing the parameterization of any arbitrary SU(4) gate \cite{vatan2004optimal}. Here, the $U_3$ gate belongs to the family of single-qubit unitary rotations. It is parameterized by three angles and can be expressed as $U_3(\theta, \phi, \lambda) = R_{z}(\phi)R_{x}(-\pi/2)R_{z}(\theta)R_{x}(\pi/2)R_{z}(\lambda)$. By choosing appropriate values for $\theta, \phi$, and $\lambda$, the state of a single qubit can be manipulated to any desired superposition state, providing the necessary flexibility for implementing convolution operations in quantum circuits. Therefore, convolutional ansatz-2 is general, allowing the QCNN to span the whole Hilbert space for any desired two-qubit unitary operation.

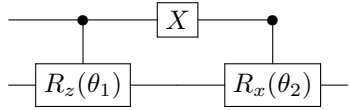
\begin{figure}[h!]
    \centering
    \resizebox{0.35\linewidth}{!}{
    \centering
    \Qcircuit @C=1em @R=1.0em { 
    & \ctrl{1} & \gate{X} & \ctrl{1} \\
    & \gate{R_z(\theta_1)} & \qw & \gate{R_x(\theta_2)} & \qw \\}
    }
    \caption{Pooling ansatz containing 2 trainable parameters that is applied on two neighboring qubits. The control qubit occupying the Pauli-$X$ gate is disregarded from further operations, and the other qubit is used in the next layer.} 
    \label{fig:2Q_pool}
\end{figure}

The pooling ansatz used in the study is shown in Fig. \ref{fig:2Q_pool}, containing two controlled rotation gates from the family $\{R_x, R_z\}$ and a Pauli-$X$ gate applied on the control qubit. The pooling layer effectively concentrates the information of two qubits into one qubit. It is done by first applying the pooling ansatz to a pair of neighboring qubits, transferring the information between qubits using the controlled rotation gates, and then the control qubit is not utilized in any further operations or measurements in the rest of the circuit.

A measurement operation is performed to obtain the classical output from the circuit. Representing the post-pooling operation state of the final 8th qubit by the density matrix $\rho_y$, the probabilities for a standard basis measurement of the qubit are obtained as

\begin{equation}
\Pr(\ket{i}) = \bra{i}\rho_y\ket{i}; \hspace{0.5cm}i=0,1
\end{equation}

Here, the probabilities of obtaining the basis \( \ket{0} \) and \( \ket{1} \) are taken from the final 8th qubit, which effectively serves as the output predictions of the model for two binary labels. Binary cross-entropy (\( L_{\mathrm{BCE}} \)) loss is utilized to measure the cross-entropy error between prediction probabilities and true class labels, expressed mathematically as

\begin{equation}
L_{\mathrm{BCE}} = - \left[ y \log \left( \Pr(\ket{1}) \right) + (1 - y) \log \left( \Pr(\ket{0}) \right) \right]
\end{equation}
where $y$ is the true label (either 0 or 1) for a data sample, $\Pr(\ket{1}) = \bra{1}\rho_y\ket{1}$ is the probability of measuring $\ket{1}$ from the final 8th qubit, and $\Pr(\ket{0}) = \bra{0}\rho_y\ket{0}$ is the probability of measuring $\ket{0}$. The overall loss is the average of individual losses across all data samples. The next sub-section describes the mechanism of the feature re-encoding layer.

\subsection{Selective Feature Re-Encoding Layer}

Since the effective qubit count decreases along the depth of the QCNN due to pooling, the input classical features at the initial encoding layer cannot be fully utilized in the re-encoding layers at intermediate stages within the circuit. However, if the classical feature vector possesses a sequential arrangement based on importance, then the most significant feature components can be selected from the vector to match the reduced number of available qubits and applied in the re-encoding operation. Principal Component Analysis (PCA) is a classical dimensionality reduction technique, facilitating this special property that is employed in the proposed method. PCA reduces the number of features while preserving the most important information in an orderly manner. It transforms the original data into orthogonal principal components arranged in a way that the first few components capture the maximum variance in the data. Let $X \in \mathbb{R}^{m \times p}$ be the original data matrix, where $m$ is the number of samples and $p$ is the number of initial data points per sample. Then, PCA involves computing the covariance matrix of $X$ followed by eigenvalue decomposition to obtain a set of eigenvectors and eigenvalues. The eigenvectors, known as principal components, are sorted in descending order based on the magnitude of corresponding eigenvalues. By selecting the first $n < p$ principal components, a reduced transformation matrix is formed and used to project the original data onto a lower-dimensional space, resulting in a transformed data matrix $X_{\text{pca}} \in \mathbb{R}^{m \times n}$. This $X_{pca}$ contains the most crucial $n$ features per sample in a sorted manner, with the first feature ($x_1$) being the highest significant and the last ($x_n$) being the least significant among them. These $n$ features are initially encoded into an $n$-qubit QCNN through angle encoding. Angle encoding \cite{schuld2021effect} is a special mechanism to transform data from the classical domain into quantum states, which uses single-qubit rotation operations to encode $n$-dimensional classical data into $n$ qubits by rotating their initial states around the Bloch sphere, with classical data points serving as the rotation parameters. Each data point $x_{i}$ is encoded in the $i$-th qubit as $\ket{\psi_{x_{i}}} = R(x_{i})\ket{\psi_{0_{i}}}$, where $\ket{\psi_{0_{i}}}$ is the initial state of $i$-th qubit and $R(\cdot) \in \{ R_{x}, R_{y}, R_{z} \}$. Thus, the overall data $x = (x_{1}, x_{2}, \cdots, x_{n})^{T}$ is encoded into $n$ qubits as:
\begin{equation}
    U_{\phi}(x): x \to \ket{\psi_{x}} = \bigotimes_{i=1}^{n} R(x_{i})\ket{\psi_{0_{i}}} = \bigotimes_{i=1}^{n}\ket{\psi_{x_{i}}}
\end{equation}
where $U_{\phi}(x)$ can be any of the rotation operations $R_{x}$, $R_{y}$ or $R_{z}$ around the respective axes of the Bloch sphere.

After reduction in the number of operational qubits through pooling, if there remain $k<n$ qubits, then the first $k$ features ($x_1 \to x_k$) are selected as the most significant ones and are re-encoded in the QCNN again through angle encoding. 
\begin{figure}[h!]
    \centering
    \includegraphics[width=0.8\linewidth]{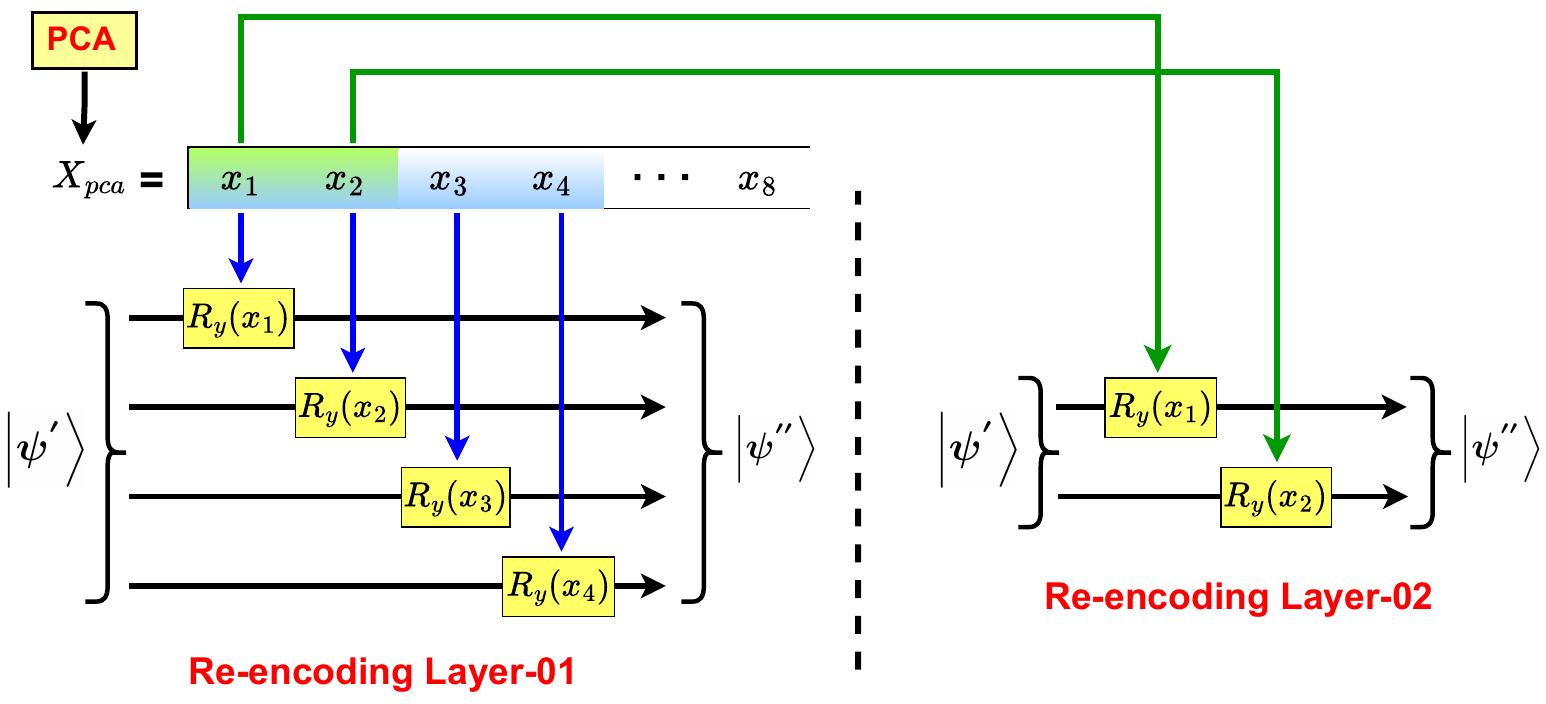}
    \caption{Structure of the proposed selective feature re-encoding layers. The feature vector extracted through PCA ($X_{pca}$) comprises $n=8$ features sorted according to significance. In the first re-encoding layer, the most significant 4 components ($x_1 \to x_4$) are encoded in the 4 remaining qubits after the first pooling. Similarly, only 2 qubits remain operational after the second pooling, and hence the most significant 2 components ($x_1$ and $x_2$) are encoded in the second re-encoding layer. For both layers, angle encoding with $R_y$ rotation is used.}
    \label{fig:reencoding_layer}
\end{figure}
In the proposed modified 8-qubit QCNN of Fig. \ref{fig:reencoded_qcnn}, all qubits are initially set in the $\ket{0}$ state. At first, $n=8$ classical features are extracted from the raw image using PCA, and all are encoded as quantum states using the angle encoding technique. The first pooling results in $\frac{n}{2} = 4$ qubits, and therefore the first 4 most significant components among the initial PCA-extracted features ($X_{pca}$) are selected for use in the first re-encoding layer. Similarly, $\frac{n}{4}=2$ qubits remain after the second pooling, and hence, the first 2 most significant features are used in the second re-encoding layer. The overall selection mechanism and the structure of the re-encoding layers used in Fig. \ref{fig:reencoded_qcnn} are illustrated in Fig. \ref{fig:reencoding_layer}. For angle encoding, the $R_y(x_i)$ rotation gate is used for both the initial encoding layer and the re-encoding layers, where the feature $x_i$ acts as the rotation angle. Mathematically, the re-encoding operation can be represented as 
\begin{align}
    \ket{\psi^{''}} = \bigotimes_{i=1}^{k} R_y(x_i)\ket{\psi^{'}_i} 
\end{align}   
where $\ket{\psi^{'}} = \otimes_{i=1}^{k} \ket{\psi_i^{'}}$ represents the quantum state after pooling, $\ket{\psi^{''}}$ is the moderated quantum state produced through the re-encoding operation, and $k$ is the number of remaining qubits after pooling. This process is inherently a unitary operation performed in the quantum domain. The moderated state $\ket{\psi^{''}}$ needs to undergo another conv-pool stage to further refine the quantum features for the desired outcome. Hence, a re-encoding layer is applied after those pooling layers, which is followed by at least one conv-pool operation, and no re-encoding is performed on the last qubit after 3rd pooling layer that does not follow any further operation except measurement. This re-encoding strategy is analogous to the self-attention mechanism used in classical neural networks, where the network is allowed to dynamically focus on important parts of the input data, enhancing its ability to capture relevant features. In this way, the re-encoding mechanism compensates for post-pooling information loss by reintroducing crucial high-variance data to redirect the focus more on the optimal solution space, thereby improving the learning and performance of QCNN.  

\section{Proposed QCNN with Joint Optimization and Ensemble Strategy}

For a QCNN model to excel in classical data classification, it must be provided with a rich set of features that are well separated between classes. Therefore, the model's performance heavily depends on the classical feature extraction technique used. However, as the combination of input data changes, the effectiveness of a single feature extraction strategy in producing features with distinguishable inter-class patterns may vary. Consequently, this can result in suboptimal features in some cases and degrade overall model performance. To address this challenge, more than one feature extraction method needs to be utilized within the QCNN framework. In this study, two of the most widely used and effective image feature extraction techniques for QCNNs, PCA and autoencoders, are considered and employed simultaneously within the QCNN framework. This dual approach ensures that if one type of feature lacks sufficient information, the other can compensate, thereby preventing the model from underperforming. 

\begin{figure}
    \centering
    \includegraphics[width=1\linewidth]{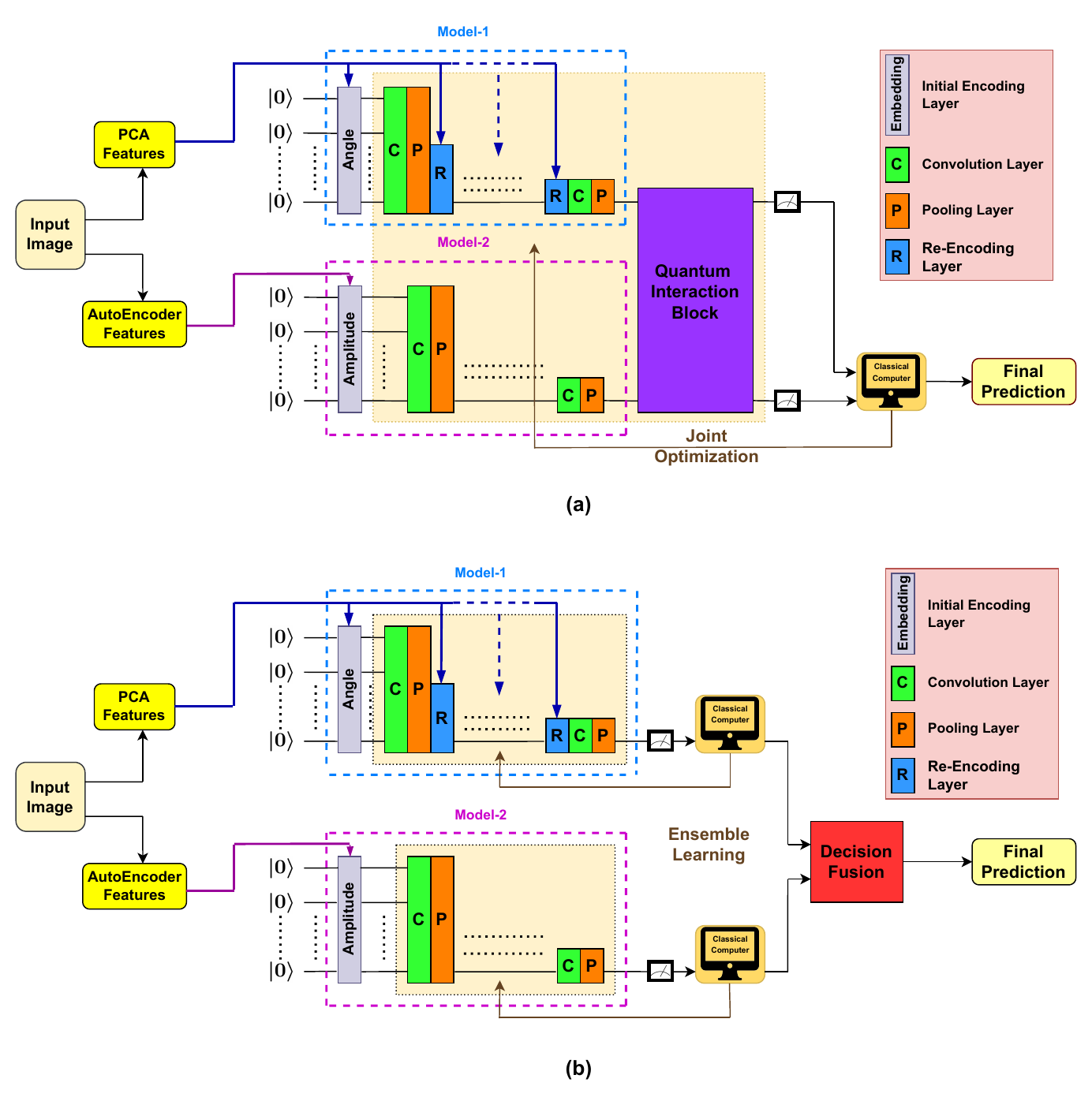}
    \caption{Framework of proposed joint optimization strategy and ensemble learning strategy, both utilizing two QCNNs (represented as Model-1 and Model-2), where Model-1 incorporates the feature re-encoding strategy. In the joint optimization approach (a), the output qubits from both Models are passed through an interaction block in the quantum domain, utilizing a common loss function and enabling a joint training operation. In contrast, the ensemble learning approach (b) independently trains each Model and performs a decision fusion technique on the output qubits' post-measurement values to obtain the final prediction.}
    \label{fig:combined_qcnn}
\end{figure}

However, the distinct feature sets derived from PCA and auto-encoder capture fundamentally different aspects of the data. PCA features emphasize capturing global variance and linear correlations, whereas the autoencoder features capture non-linear and learned latent representations. Processing these heterogeneous feature sets directly into a single QCNN could potentially blend their unique representational strengths, obscuring the unique correlations inherent in each type of feature. Hence, a single QCNN circuit might struggle to optimally encode and process such diverse properties within a single Hilbert space. In this study, two separate QCNNs are employed to process the two distinct feature sets, with their output stages integrated to produce the final prediction. Using separate QCNNs allows each network to specialize in exploiting the intrinsic properties of its corresponding feature type. Rather than feeding both feature types within a single QCNN framework, this approach is designed to maximize the efficacy of different feature representations and model performance. 

Initially, two different QCNN models are constructed with distinct encoding methods, each utilizing specific classical features (PCA or autoencoder features), to judge their independent contributions in the classification task. These QCNNs are referred to as Model-1 and Model-2, respectively. To consolidate the information acquired from these models and leverage their strengths, two additional QCNN frameworks are developed by integrating the output stages of Model-1 and Model-2. In the first approach, a quantum domain interaction is established between the circuits of Model-1 and Model-2, where the overall structure is jointly trained with a single shared loss function and provides the final prediction. This framework, employing the joint optimization strategy, is referred to as Model-3. In the second approach, Model-1 and Model-2 are independently trained, and their output scores are combined as an ensemble to make the final prediction. This framework, employing the ensemble learning strategy, is referred to as Model-4. The general schematic architecture of joint optimization and ensemble learning methods is demonstrated in Fig. \ref{fig:combined_qcnn}. The joint optimization approach (Model-3) allows mutual adjustment of parameters during training and enhances the collaborative learning process, ensuring that the adjustment of trainable weights in one QCNN would depend on the changes in the quantum state orientations of the other QCNN. On the other hand, Model-4 utilizes a simple ensemble learning mechanism with individual training of component models, and then combines their post-measurement output values through decision fusion. This is implemented to validate the effectiveness of the unified training approach used in Model-3 over the independent training scheme used in Model-4. Details of the proposed QCNN frameworks are described in the following sub-sections.    

\subsection{Model-1: QCNN with PCA Features}\label{sec4_sub1}

This model is the 8-qubit QCNN shown in Fig. \ref{fig:reencoded_qcnn}, utilizing PCA features for initial encoding and subsequent re-encoding via the angle encoding mechanism. The encoding mechanism is chosen such that it better aligns with the inherent statistical characteristics of the features. PCA is a linear transformation technique that emphasizes variance maximization. The extracted principal components have an ordered and structured representation (ranked by importance). Angle encoding is particularly suitable for such data, as it maps each feature component to quantum rotation angles, thus effectively preserving relative variations. The structural configuration of Model-1 (Fig. \ref{fig:reencoded_qcnn}) is explained in earlier sections. It contains two extra feature re-encoding layers within the conventional QCNN framework. PCA reduces the raw image to an 8-component feature vector. These features are initially encoded, and selective components are subsequently re-encoded at later stages using the angle encoding technique. After consecutive conv-pool operations, the probability measurements are taken from the final qubit, which serve as the model's prediction for binary classification. The BCE loss ($L_{BCE}$) is calculated based on the output probabilities and the true labels.

\subsection{Model-2: QCNN with Autoencoder Features}\label{sec4_sub2}

The second QCNN model follows the conventional structure, consisting of only 4 qubits. At first, a classical auto-encoder is used to extract a feature vector of length 16 from the raw image, which is encoded using amplitude encoding. In the amplitude encoding mechanism \cite{larose2020robust}, the classical data is mapped into the normalized probability amplitudes of quantum computational basis states using the quantum superposition principle. For $n$-qubit system, amplitude encoding can transform $N = 2^n$ dimensional classical data $x = (x_{1}, x_{2}, \cdots, x_{N})^{T}$ into the probability amplitudes of $N$ possible computational basis states as
\begin{equation}
     U_{\phi}(x): x \to \ket{\psi_{x}} = \frac{1}{\left\| x \right\|}\sum_{i=1}^{N} x_{i} \ket{i}
     \label{eq:ae}
\end{equation}
where $x_{i}$ is encoded as the probability amplitude of the $i$-th computational basis state $\ket{i}$, and $\left\| x \right\|$ is the normalization factor.  

\begin{figure}
    \centering
    \includegraphics[width=0.65\linewidth]{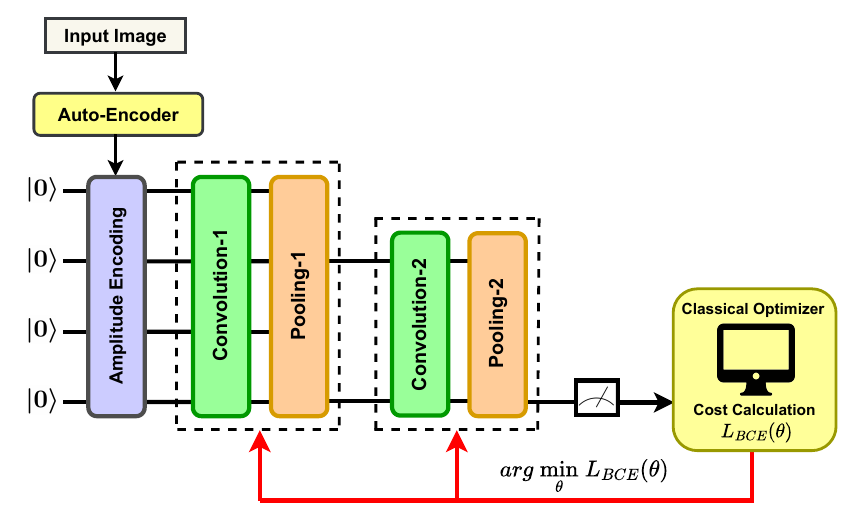}
    \caption{Architecture of Model-2. The auto-encoder derived features are encoded into 4 qubits using the amplitude encoding method. After two stages of conv-pool operations, a measurement is taken for the cost calculation and parameter update.}
    \label{fig:4_qcnn}
\end{figure}

Autoencoder features capture complex and non-linear latent structures, which usually have more intricate distributions, with potentially richer and subtler amplitude relationships among the features. Amplitude encoding is highly suitable in this case, as it efficiently maps classical data directly into the amplitudes of quantum basis states, effectively capturing nonlinear and multidimensional feature relationships. It also allows for a large number of complex features to be represented with a smaller number of qubits (16 features encoded into $log_216=4$ qubits). Being a smaller circuit than Model-1, Model-2 comprises only two conv-pool operations, after which the probabilistic measurement is taken from the last remaining qubit that provides class prediction for the respective input data. The loss function is the same as that used for Model-1. The architecture is illustrated in Fig. \ref{fig:4_qcnn}.

\subsection{Joint Optimization of Two QCNNs} \label{sec4_sub2_sub3} 

While combining the learning outcomes of separate models, the joint training approach offers several advantages over the independent training scheme. Here, the interaction of the component systems is taking place within the quantum feature domain while the model is still in its training phase. This is a quantum equivalent of the feature-level fusion used in classical deep learning models. As individual component systems are fed with different feature types, the joint optimization technique facilitates the overall feature processing in a more coordinated way and achieves an effective balance of these features by complementing each other. More specifically, the complex quantum interactions between processed PCA features and autoencoder features can produce richer information in this approach that would not be apparent from the independent training scheme. Thus, end-to-end training with a single loss function ensures consolidated focus solely on the final objective, thereby reducing the risk of overfitting and obtaining better generalization.

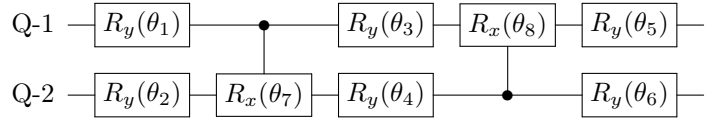
\begin{figure}[h!]
    \begin{center}
        \begin{minipage}{0.65\linewidth}
            \centering
            \Qcircuit @C=1em @R=1em {
                \lstick{\text{Q-1}} & \gate{R_y(\theta_1)} & \ctrl{1} & \gate{R_y(\theta_3)} & \gate{R_x(\theta_8)} & \gate{R_y(\theta_5)} & \qw \\
                \lstick{\text{Q-2}} & \gate{R_y(\theta_2)} & \gate{R_x(\theta_7)} & \gate{R_y(\theta_4)} & \ctrl{-1} & \gate{R_y(\theta_6)} & \qw \\
            }
        \end{minipage}
    \end{center}
    \caption{Proposed architecture of the Interaction Block, which implements quantum interaction between the output qubits from Model-1 (Q-1) and Model-2 (Q-2). It consists of 6 $R_y$ rotation gates and 2 controlled $R_x$ rotation gates, with a total of 8 trainable parameters.}
    \label{fig:interaction_block}
\end{figure}

The proposed joint optimization approach of two QCNN models is depicted in Fig. \ref{fig:combined_qcnn}(a). The last qubit from Model-1 is taken after three consecutive conv-pool operations with re-encoding layers in between. Similarly, the last qubit from Model-2 is obtained after two conv-pool stages. These two resultant qubits are passed through a quantum interaction block, which is designed to facilitate quantum domain interaction between the two models. The circuit configuration of the interaction block is shown in Fig. \ref{fig:interaction_block}. It contains 6 $R_{y}$ gates and 2 controlled $R_{x}$ gates, comprising a total of 8 trainable parameters. The first controlled rotation gate enables the qubit from Model-1 (control) to influence the qubit from Model-2 (target). This operation entangles the qubits, allowing them to interact in a way that leverages the encoded information from both QCNN networks. Then, the reversed controlled rotation gate allows the qubit from Model-2 (control) to influence the qubit from Model-1 (target). Therefore, a bidirectional interaction occurs between the two qubits. Through these operations, the interaction block facilitates an effective quantum interaction between the outputs of the component models. 

Here, Model-1 and Model-2 are undergoing the interaction block while being in quantum states, creating a coupled optimization process. The parameter tuning for one Model will depend on the quantum state orientation of the other Model. It guarantees that the optimization process considers the combined effect of both Models. 
Following the interaction block, which facilitates mutual information transfer between the sequentially processed two qubits focusing on different classical features, both final qubits are considered for measurement operation rather than arbitrarily selecting one. Measuring a single qubit might neglect critical entangled information from its counterpart, whereas assessing both qubits ensures the loss function optimizes the complete interdependent quantum state.

For measurement, the Pauli-\( Z \) expectation values are taken from the final two qubits, followed by a softmax transformation to derive class-wise prediction probabilities, which are required for calculating the cross-entropy loss function. The expectation value \( \expval{Z} \) of an observable operator $\hat{Z} \otimes \hat{Z}$ in the final reduced 2-qubit  state \( \rho_y \) can be mathematically expressed as:

\begin{equation}
\expval{ZZ} = \Tr(\rho_y (\hat{Z}\otimes\hat{Z}))
\end{equation}

Here, the linear operator on the computational basis is represented by the Hermitian matrix \( Z \). The range of measured Pauli-\( Z \) expectation values lies within \( [-1, 1] \). A mapping from expectation values from the two qubits to probability values for binary classes is performed using the softmax function defined as:

\begin{equation}
P(y = i \mid \vb{z}) = \frac{e^{z_i}}{e^{z_1} + e^{z_2}} \quad \text{for } i = 1, 2
\end{equation}

where \( P(y = i \mid \vb{z}) \) is the probability of the input being classified as class \( i \), and \( z_i = \Tr(\rho_y \hat{Z}_i) \) is the Pauli-\( Z \) expectation value corresponding to class \( i \). In the context of this model, \( z_1 \) and \( z_2 \) are the expectation values obtained from the final two qubits. The softmax function ensures that these probabilities sum to 1. Then, the cross-entropy loss function is calculated for performing gradient descent and updating the model parameters.

\subsection{Ensemble Strategy of Two QCNNs}\label{sec4_sub2_sub4}

In this subsection, the ensemble strategy of two independent QCNN models, referred to as Model-4, is discussed. Unlike Model-3, where the QCNN models were interacting within the quantum domain, Model-4 individually trains two QCNN circuits (Model-1 and Model-2), each with a separate loss function calculation and parameter update. The overall workflow is depicted in Fig. {\ref{fig:combined_qcnn}}(b). The final prediction in Model-4 is made by taking a sum-rule decision fusion of the individual post-measurement output values from each QCNN.

After the consecutive conv-pool operations, each QCNN circuit reduces to a single operational qubit. The probabilistic measurement of these qubits into basis states (\(|0\rangle\) and \(|1\rangle\)) yields a prediction vector for each circuit, representing the probabilities of the input data belonging to class-1 and class-2, respectively. For a single input data sample, both QCNN circuits provide a prediction vector of two components. Mathematically, let \( \hat{y}_1 \) and \( \hat{y}_2 \) represent the prediction vectors from Model-1 and Model-2, respectively. The ensemble prediction \(\hat{y}\) using the sum-rule decision fusion method is given by

\begin{equation}
    \hat{y} = \hat{y}_1 + \hat{y}_2
\end{equation}
Here, the ensemble prediction equally prioritizes each prediction from the models. The final predicted class is determined by taking the argmax of the summed prediction vector
\begin{equation}
    \hat{c} = \arg\max(\hat{y})
\end{equation}
This method leverages the combined output of both QCNN models without any need for weight-tuning, lowering the computational expense while still benefiting from the diverse feature representations captured by each model.

\section{Simulation Results and Analysis}
Extensive experiments have been performed using multiple datasets to demonstrate the effectiveness of the approaches proposed in this study. These experiments involve different combinations of ansatzes and models, and the outcomes are comprehensively analyzed for comparison between approaches. The following subsections provide details on the datasets used, the experimental setup, and the thorough analysis of experimental outcomes.

\subsection{Datasets}
Throughout the study, two widely used standard public datasets, MNIST \cite{deng2012mnist} and Fashion MNIST \cite{xiao2017fashion}, have been utilized for experimentation. The MNIST dataset consists of grayscale images of handwritten digits (0-9) divided into 10 classes, whereas the Fashion MNIST dataset contains grayscale images of different clothing items with 10 classes. Both datasets comprise 60,000 training images and 10,000 test images, each having a $28\times28$ resolution. For investigating binary classification, three pairs of classes are selected from each dataset. From MNIST, the considered pairs are 0 vs 1, 1 vs 2, and 2 vs 3, and from the Fashion MNIST, the considered pairs are T-shirt vs Trouser, Trouser vs Dress, and Sandal vs Ankle Boot. The distribution of training and test images for each class pair from both datasets is presented in Table \ref{tab:dataset}.

\begin{table}[h!]
\centering
\caption{Distribution of training and test images for each binary class pair from both datasets.}
\begin{tabular}{lll} 
\hline
\multicolumn{3}{l}{\textbf{MNIST}}                           \\ 
\textbf{Binary Class Pair}   & \textbf{Train} & \textbf{Test}  \\ 
\hline
0 vs 1                      & 12,665         & 2115           \\
1 vs 2                       & 12,700         & 2167           \\
2 vs 3                       & 12,089         & 2042           \\ 
\hline
\multicolumn{3}{l}{\textbf{Fashion MNIST}}                   \\ 
\textbf{Binary Class Pair}   & \textbf{Train} & \textbf{Test}  \\ 
\hline
T-Shirt vs Trouser   & 12,000         & 2000           \\
Trouser vs Dress     & 12,000         & 2000           \\
Sandal vs Ankle Boot & 12,000         & 2000           \\
\hline
\end{tabular}
\label{tab:dataset}
\end{table}

\subsection{Experimental Setup}

All experiments in this study are done using two convolutional ansatzes, referred to as Convolutional Ansatz 1 (Fig. \ref{fig:SO4_ansatz}), Convolutional Ansatz 2 (Fig. \ref{fig:SU(4)}), and a quantum pooling ansatz (Fig. \ref{fig:2Q_pool}). Properties of these ansatzes have been covered earlier in Section \ref{feature_reencod}. 

For image dimensionality reduction, PCA and autoencoder methods are employed. The \textit{sklearn.decomposition.PCA} module from the scikit-learn library \cite{scikit-learn} has been used for PCA feature extraction. During auto-encoder training, minimal over-fitting is crucial, and computational resource usage should be minimized since it is a pre-processing step. That is why a simple autoencoder with one hidden layer having a latent space of 16 dimensions is used, trained for a single epoch only. For optimizing the trainable parameters of the QCNN models, the Nesterov Momentum Optimization algorithm \cite{nesterov1983method} is employed to minimize the binary cross-entropy loss function. During the training, a mini-batch of size 25 has been randomly selected on each iteration. The total number of iterations has been set to 200. Training on the mini-batch assists in the gradients' escape from local minima. To assess the effectiveness of the proposed models and evaluate their performance, some widely recognized metrics, namely accuracy, precision, recall, and F1 score, have been used. Each of the proposed model combinations has been tested for 5 independent runs with different random initializations, and the mean scores for each metric are reported.

\subsection{Effect of the Feature Re-Encoding Strategy on QCNN}
The impact of the proposed selective feature re-encoding strategy on the QCNN performance has been analyzed across multiple binary combinations from the MNIST and Fashion MNIST datasets. For each case, the performance of the conventional QCNN has been compared with the feature re-encoded QCNN, where each QCNN architecture was explored with two different convolutional ansatzes. 

\begin{figure*}[h!]
    \centering
    \includegraphics[width=0.49\linewidth]{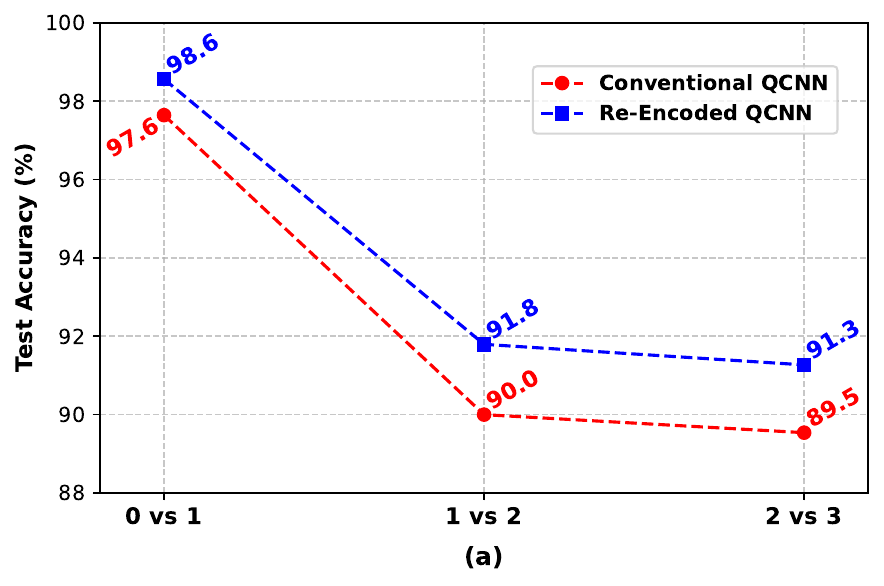} 
    \includegraphics[width=0.49\linewidth]{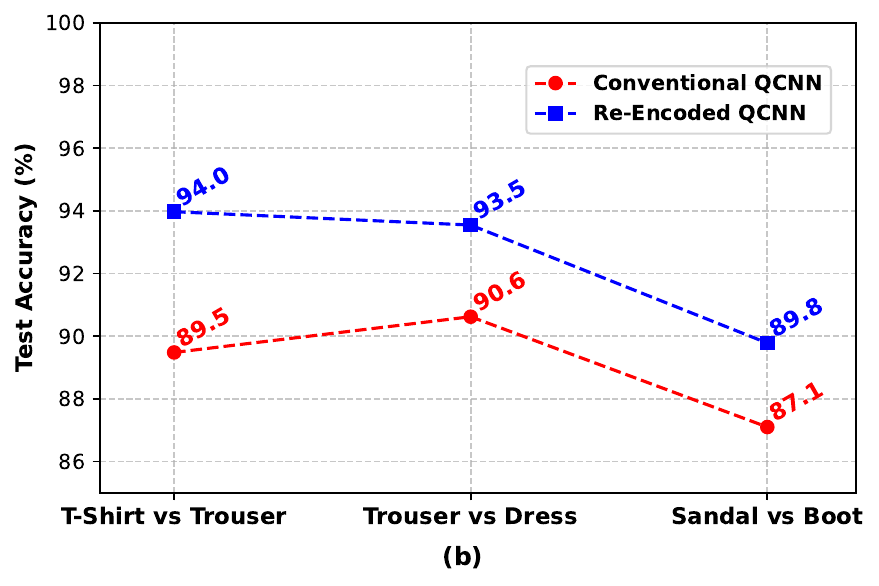}
    \caption{Average Accuracy Improvement of Feature Re-Encoded QCNN over Conventional QCNN: (a) MNIST Dataset and (b) Fashion MNIST Dataset.}
    \label{fig:compare1}
\end{figure*}

The classification results for all four metrics on the MNIST and Fashion MNIST datasets are presented in Table \ref{tab:re-embedd_m} and Table \ref{tab:fm_re-embedd}, respectively. It is evident that employing the feature re-encoding strategy led to enhanced accuracy for all binary pairs in the datasets, with improvements in other evaluation metrics as well. However, a slight decline is observed in some pairs, especially in precision for MNIST, and in recall for Fashion MNIST. The average accuracy comparison considering the two ansatzes between the feature re-encoded QCNN and the conventional architecture is illustrated in the line graph of Fig. \ref{fig:compare1}. For the MNIST dataset, the average performance improved by 1\% for the `0 vs 1' case and 1.8\% for both the `1 vs 2' and `2 vs 3' cases. On the other hand, the improvement margin is comparatively more prominent for the Fashion MNIST dataset. Since Fashion MNIST is a more complicated dataset than MNIST, the findings indicate that, when input data complexity increases, the margin of important information loss as quantum states get processed along the circuit also increases, therefore re-encoding significant feature components back into the circuit to guide the computation toward an optimal solution becomes more effective.

\begin{table*}[h!]
\centering
\caption{Performance of QCNN with and without the Re-Encoding strategy on the MNIST Dataset.}
\label{tab:re-embedd_m}
\resizebox{\textwidth}{!}{%
\begin{tabular}{ccccccc} 
\hline
\textbf{Binary~Class} & \textbf{Conv~Ansatz} & \textbf{QCNN Type} & \textbf{Accuracy} & \textbf{Precision} & \textbf{Recall} & \textbf{F1 Score} \\ 
\toprule
\multirow{4}{*}{0 vs 1} & \multirow{2}{*}{Ansatz-1} & Conventional & 0.9761 & \textbf{0.9811} & 0.9612 & 0.9738 \\
 &  & Re-Encoded & \textbf{0.9830} & 0.9805 & \textbf{0.9850} & \textbf{0.9837} \\
 & \multirow{2}{*}{Ansatz-2} & Conventional & 0.9768 & 0.9839 & 0.9657 & 0.9747 \\
 &  & Re-Encoded & \textbf{0.9883} & \textbf{0.9867} & \textbf{0.9887} & \textbf{0.9872} \\ 
\midrule
\multirow{4}{*}{1 vs 2} & \multirow{2}{*}{Ansatz-1} & Conventional & 0.8965 & 0.8480 & 0.9705 & 0.9085 \\
 &  & Re-Encoded & \textbf{0.9166} & \textbf{0.8794} & \textbf{0.9715} & \textbf{0.9255} \\
 & \multirow{2}{*}{Ansatz-2} & Conventional & 0.9035 & 0.8561 & 0.9806 & 0.9141 \\
 &  & Re-Encoded & \textbf{0.9193} & \textbf{0.8785} & \textbf{0.9835} & \textbf{0.9292} \\ 
\midrule
\multirow{4}{*}{2 vs 3} & \multirow{2}{*}{Ansatz-1} & Conventional & 0.8854 & 0.9345 & 0.8081 & 0.8669 \\
 &  & Re-Encoded & \textbf{0.9094} & \textbf{0.9393} & \textbf{0.8895} & \textbf{0.9117} \\
 & \multirow{2}{*}{Ansatz-2} & Conventional & 0.9054 & \textbf{0.9323} & 0.8817 & 0.9058 \\
 &  & Re-Encoded & \textbf{0.9160} & 0.9285 & \textbf{0.9061} & \textbf{0.9178} \\
\bottomrule
\end{tabular}
}
\end{table*}

\begin{table*}[h!]
\centering
\caption{Performance of QCNN with and without the Re-Encoding strategy on the Fashion MNIST Dataset.}
\label{tab:fm_re-embedd}
\resizebox{\textwidth}{!}{
\begin{tabular}{ccccccc} 
\hline
\textbf{Binary~Class} & \textbf{Conv~Ansatz} & \textbf{QCNN Type} & \textbf{Accuracy} & \textbf{Precision} & \textbf{Recall} & \textbf{F1 Score} \\ 
\toprule
\multirow{4}{*}{T-Shirt vs Trouser} & \multirow{2}{*}{Ansatz-1} & Conventional & 0.8882 & 0.8426 & 0.9560 & 0.8955 \\
 &  & Re-Encoded & \textbf{0.9335} & \textbf{0.9057} & \textbf{0.9693} & \textbf{0.9358} \\
 & \multirow{2}{*}{Ansatz-2} & Conventional & 0.9015 & 0.8446 & 0.9820 & 0.9090 \\
 &  & Re-Encoded & \textbf{0.9460} & \textbf{0.9160} & \textbf{0.9880} & \textbf{0.9483} \\ 
\midrule
\multirow{4}{*}{Trouser vs Dress} & \multirow{2}{*}{Ansatz-1} & Conventional & 0.9015 & 0.9267 & \textbf{0.8733} & 0.8958 \\
 &  & Re-Encoded & \textbf{0.9190} & \textbf{0.9405} & 0.8690 & \textbf{0.9064} \\
 & \multirow{2}{*}{Ansatz-2} & Conventional & 0.9110 & 0.9400 & 0.8780 & 0.9079 \\
 &  & Re-Encoded & \textbf{0.9520} & \textbf{0.9704} & \textbf{0.9330} & \textbf{0.9495} \\ 
\midrule
\multirow{4}{*}{Sandal vs Ankle Boot} & \multirow{2}{*}{Ansatz-1} & Conventional & 0.8731 & 0.8836 & \textbf{0.8506} & 0.8620 \\
 &  & Re-Encoded & \textbf{0.8997} & \textbf{0.9310} & 0.8460 & \textbf{0.8842} \\
 & \multirow{2}{*}{Ansatz-2} & Conventional & 0.8690 & 0.8884 & 0.8440 & 0.8656 \\
 &  & Re-Encoded & \textbf{0.8960} & \textbf{0.9250} & \textbf{0.8620} & \textbf{0.8923} \\
\bottomrule
\end{tabular}
}
\end{table*}

\subsection{Performance of Joint Optimization and Ensemble Strategy}

In this subsection, the performance evaluation of the four QCNN models developed for the joint optimization approach and ensemble strategy is performed across different binary combinations of the MNIST and Fashion MNIST datasets. The models are Model-1 (QCNN with PCA features), Model-2 (QCNN with autoencoder features), Model-3 (integration of Model-1 and Model-2 with quantum interaction block and joint optimization), and Model-4 (decision fusion-based ensemble technique of Model-1 and Model-2).


\begin{table*}[h!]
\centering
\caption{Performance of different QCNN Models on the MNIST dataset.}
\label{tab:table_Performance_MNIST}
\resizebox{\textwidth}{!}{
\begin{tabular}{ccccccc} 
\hline
\textbf{Binary Class} & \textbf{Conv Ansatz} & \textbf{QCNN Model} & \textbf{Accuracy (\%)} & \textbf{Precision} & \textbf{Recall} & \textbf{F1 Score} \\ 
\toprule
\multirow{8}{*}{0 vs 1} & \multirow{4}{*}{Ansatz-1} & Model-1 & 98.10 & 0.9795 & 0.9830 & 0.9817 \\
 &  & Model-2 & \textbf{99.70} & \textbf{0.9989} & 0.9939 & \textbf{0.9964} \\
 &  & Model-3 & 99.58 & 0.9959 & \textbf{0.9948} & 0.9953 \\
 &  & Model-4 & 99.38 & 0.9943 & 0.9922 & 0.9933 \\ 
\cmidrule{2-7}
 & \multirow{4}{*}{Ansatz-2} & Model-1 & 98.63 & 0.9847 & 0.9857 & 0.9852 \\
 &  & Model-2 & \textbf{99.81} & \textbf{0.9970} & \textbf{0.9989} & \textbf{0.9979} \\
 &  & Model-3 & 99.74 & 0.9967 & 0.9977 & 0.9972 \\
 &  & Model-4 & 99.62 & 0.9949 & 0.9969 & 0.9960 \\ 
\midrule
\multirow{8}{*}{1 vs 2} & \multirow{4}{*}{Ansatz-1} & Model-1 & 91.46 & 0.8794 & 0.9700 & 0.9225 \\
 &  & Model-2 & 96.06 & 0.9552 & 0.9730 & 0.9643 \\
 &  & Model-3 & \textbf{97.44} & \textbf{0.9633} & 0.9888 & \textbf{0.9759} \\
 &  & Model-4 & 96.20 & 0.9387 & \textbf{0.9919} & 0.9643 \\ 
\cmidrule{2-7}
 & \multirow{4}{*}{Ansatz-2} & Model-1 & 91.93 & 0.8785 & 0.9815 & 0.9272 \\
 &  & Model-2 & 96.02 & 0.9392 & 0.9861 & 0.9628 \\
 &  & Model-3 & \textbf{97.35} & \textbf{0.9598} & 0.9920 & \textbf{0.9760} \\
 &  & Model-4 & 95.85 & 0.9279 & \textbf{0.9972} & 0.9611 \\ 
\midrule
\multirow{8}{*}{2 vs 3} & \multirow{4}{*}{Ansatz-1} & Model-1 & 90.74 & 0.9245 & 0.8895 & 0.9067 \\
 &  & Model-2 & 90.75 & 0.9367 & 0.8760 & 0.9054 \\
 &  & Model-3 & \textbf{92.77} & 0.9382 & \textbf{0.9176} & \textbf{0.9276} \\
 &  & Model-4 & 91.72 & \textbf{0.9430} & 0.8905 & 0.9155 \\ 
\cmidrule{2-7}
 & \multirow{4}{*}{Ansatz-2} & Model-1 & 91.41 & 0.9245 & 0.9041 & 0.9138 \\
 &  & Model-2 & 91.04 & \textbf{0.9648} & 0.8542 & 0.9056 \\
 &  & Model-3 & \textbf{93.43} & 0.9490 & \textbf{0.9189} & \textbf{0.9335} \\
 &  & Model-4 & 92.63 & 0.9425 & 0.9095 & 0.9257 \\
\bottomrule
\end{tabular}
}
\end{table*}

\begin{figure}[h!]
    \centering
    \includegraphics[width=1\linewidth]{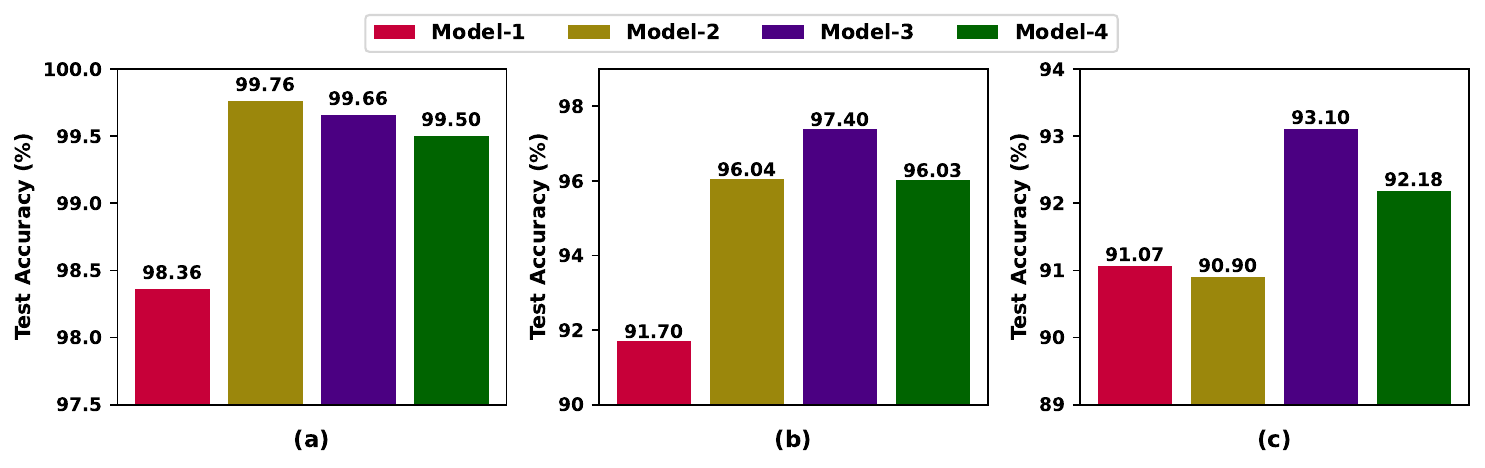}
    \caption{Comparison of average accuracy across different Models on the MNIST dataset, (a) 0 vs 1, (b) 1 vs 2, and (c) 2 vs 3.}
    \label{fig:joint_mnist}
\end{figure}

\begin{figure}[h!]
    \centering
    \includegraphics[width=1\linewidth]{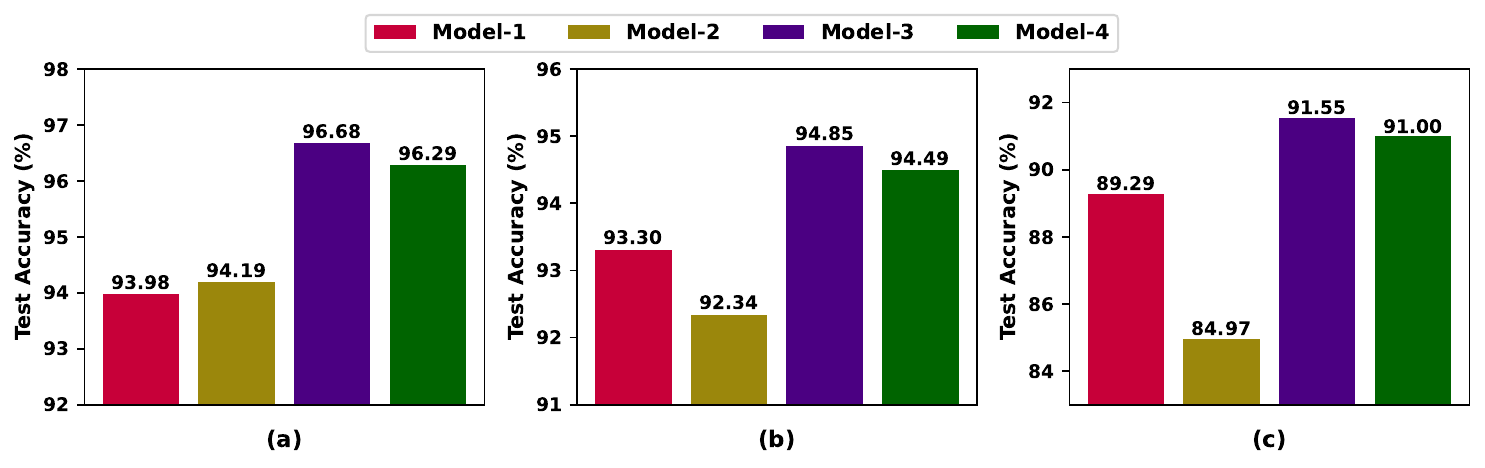}
    \caption{Comparison of average accuracy across different Models on the Fashion MNIST dataset, (a) T-Shirt vs Trouser, (b) Trouser vs Dress, and (c) Sandal vs Ankle Boot.}
    \label{fig:joint_f_mnist}
\end{figure}

The performance of different QCNN models on the standard metrics across three binary classification tasks from the MNIST and Fashion MNIST datasets is presented in Table \ref{tab:table_Performance_MNIST} and \ref{tab:table_performance_FashionMNIST} respectively. The corresponding average accuracy comparison is demonstrated in the bar charts of Fig. \ref{fig:joint_mnist} and Fig. \ref{fig:joint_f_mnist} as well. The results demonstrate that Model-3 and Model-4, which integrate two quantum circuits encoded with different classical features, exhibit superior performance compared to the individual models (Model-1 and Model-2). Model-3 consistently achieves the highest accuracy, whereas Model-4 remains slightly behind Model-3. An exception emerges in the MNIST 0 vs. 1 classification, where Model-2 demonstrates the maximum accuracy under both ansatzes, slightly edging out Model-3 and Model-4. This anomaly can be attributed to the near-perfect separability of this digit pair, leading to a saturation effect where all models approach complete accuracy. In particular, Model-2, with autoencoder features, excels in this saturated scenario, leaving minimal room for further improvement by joint optimization or ensemble techniques. For Fashion MNIST dataset, no saturation effect is observed, since this dataset has greater complexity and variability. As a result, the individual models are not dominating due to exceptional feature separability. Rather, the stable superiority of the accuracy measure in Model-3 and Model-4 affirms the benefit of combining distinct QCNN circuits encoded with distinct classical features.

\begin{table*}[h!]
\centering
\caption{Performance of different QCNN Models on the Fashion MNIST dataset.}
\label{tab:table_performance_FashionMNIST}
\resizebox{\textwidth}{!}{
\begin{tabular}{c c c c c c c}
\toprule
\textbf{Classes} & \textbf{Conv.} & \textbf{QCNN} & \textbf{Accuracy (\%)} & \textbf{Precision} & \textbf{Recall} & \textbf{F1 Score} \\
\midrule
\multirow{8}{*}{T-Shirt vs Trouser} & \multirow{4}{*}{Ansatz-1} & Model-1 & 93.35 & 0.9057 & 0.9683 & 0.9358 \\
 &  & Model-2 & 93.95 & 0.9128 & 0.9723 & 0.9415 \\
 &  & Model-3 & \textbf{96.85} & \textbf{0.9580} & 0.9800 & \textbf{0.9688} \\
 &  & Model-4 & 96.37 & 0.9447 & \textbf{0.9850} & 0.9644 \\
\cmidrule{2-7}
 & \multirow{4}{*}{Ansatz-2} & Model-1 & 94.60 & 0.9160 & 0.9830 & 0.9483 \\
 &  & Model-2 & 94.42 & 0.9361 & 0.9540 & 0.9450 \\
 &  & Model-3 & \textbf{96.50} & \textbf{0.9697} & 0.9600 & \textbf{0.9643} \\
 &  & Model-4 & 96.20 & 0.9410 & \textbf{0.9860} & 0.9628 \\
\midrule
\multirow{8}{*}{Trouser vs Dress} & \multirow{4}{*}{Ansatz-1} & Model-1 & 91.50 & 0.9405 & 0.8590 & 0.9014 \\
 &  & Model-2 & 92.65 & 0.9670 & 0.8830 & 0.9230 \\
 &  & Model-3 & \textbf{93.73} & 0.9529 & \textbf{0.9210} & \textbf{0.9364} \\
 &  & Model-4 & 93.15 & \textbf{0.9840} & 0.8780 & 0.9270 \\
\cmidrule{2-7}
 & \multirow{4}{*}{Ansatz-2} & Model-1 & 95.10 & 0.9704 & 0.9330 & 0.9495 \\
 &  & Model-2 & 92.02 & 0.9315 & 0.9130 & 0.9200 \\
 &  & Model-3 & \textbf{95.96} & \textbf{0.9834} & 0.9378 & \textbf{0.9600} \\
 &  & Model-4 & 95.82 & 0.9706 & \textbf{0.9472} & 0.9576 \\
\midrule
\multirow{8}{*}{Sandal vs Ankle Boot} & \multirow{4}{*}{Ansatz-1} & Model-1 & 88.97 & 0.9310 & 0.8420 & 0.8842 \\
 &  & Model-2 & 82.13 & 0.9025 & 0.7220 & 0.8015 \\
 &  & Model-3 & \textbf{91.55} & \textbf{0.9738} & 0.8540 & \textbf{0.9099} \\
 &  & Model-4 & 90.80 & 0.9473 & \textbf{0.8640} & 0.9038 \\
\cmidrule{2-7}
 & \multirow{4}{*}{Ansatz-2} & Model-1 & 89.60 & 0.9250 & 0.8620 & 0.8923 \\
 &  & Model-2 & 87.80 & \textbf{0.9385} & 0.8090 & 0.8689 \\
 &  & Model-3 & \textbf{91.55} & 0.9152 & \textbf{0.9070} & \textbf{0.9110} \\
 &  & Model-4 & 91.20 & 0.9310 & 0.8900 & 0.9091 \\
\bottomrule
\end{tabular}
}
\end{table*}

\subsection{Discussion}

By integrating the individual QCNN models, both the joint optimization and the ensemble learning approaches demonstrate satisfactory performance, improving generalization capability. However, the comparative analysis highlights the superiority of the joint optimization approach across all considered binary classification combinations of the datasets over ensemble learning. Since this method integrates two QCNN circuits through a quantum interaction block, which facilitates inter-network entanglement and coherent parameter optimization across the networks, it enhances the networks' ability to represent complex features more effectively. In contrast, the ensemble method lacks this unified training synergy, leading to suboptimal coordination between the networks, as their outputs are fused post-training rather than co-optimized. Another observation is the performance disparity between the two convolutional ansatzes. Ansatz-2 consistently outperforms Ansatz-1 across both datasets, attributable to its greater number of trainable parameters, which enhances representational capacity. For the proposed best-performing joint optimization method, the average accuracy on the MNIST dataset across the three binary tasks is 96.60\% for Ansatz-1 and 96.84\% for Ansatz-2, yielding a 0.24\% improvement margin for Ansatz-2. Similarly, for Fashion MNIST, the corresponding average accuracy is 94.04\% for Ansatz-1 and 94.67\% for Ansatz-2, reflecting a 0.63\% increase. Though this improvement margin is minor, it still underscores Ansatz-2’s advantage due to increased parameter flexibility.

A comparison between state-of-the-art works of binary classifications on MNIST, Fashion MNIST datasets, and the proposed QCNN models is presented in Table \ref{table_compare_mnist_bench}. The comparison clearly shows that the proposed models persistently outperform the accuracy achieved by other benchmark methods across the specified datasets.

\begin{table*}[h]
\caption{Comparison of the Proposed Model with the existing methods.}
\label{table_compare_mnist_bench}
\centering
\resizebox{0.75\columnwidth}{!}{%
\begin{tabular}{c l c}
\toprule
\textbf{Dataset} & \textbf{Model Used} & \textbf{Accuracy (\%)} \\
\midrule
\multirow{5}{*}{\shortstack{Fashion MNIST\\ (T-Shirt vs Trouser)}} & Easom et al. \cite{easom2022efficient} & 89.50 \\
 & Hur et al. \cite{hur2022quantum} & 94.30 \\
 & Mahmud et al. \cite{mahmud2024quantum} & 95.75 \\
 & Model-4 (Ensemble Strategy) & 96.20 \\
 & \textbf{Proposed Model-3 (Joint Opt.)} & \textbf{96.50} \\
\midrule
\multirow{6}{*}{MNIST (0 vs 1)} & Easom et al. \cite{easom2022efficient} & 94.60 \\
 & Kim et al. \cite{kim2023classical} & 98.50 \\
 & Hur et al. \cite{hur2022quantum} & 98.70 \\
 & Mahmud et al. \cite{mahmud2024quantum} & 99.00 \\
 & Model-4 (Ensemble Strategy) & 99.62 \\
 & \textbf{Proposed Model-3 (Joint Opt.)} & \textbf{99.74} \\
\midrule
\multirow{3}{*}{MNIST (2 vs 3)} & Kim et al. \cite{kim2023classical} & 90.00 \\
 & Model-4 (Ensemble Strategy) & 92.63 \\
 & \textbf{Proposed Model-3 (Joint Opt.)} & \textbf{93.43} \\ 
\bottomrule
\end{tabular}%
}
\end{table*}

\section{Conclusion}

In this paper, an efficient QML architecture aimed at image classification is developed by jointly optimizing two separate QCNN models, where one model utilizes PCA-based data reduction with the proposed classical feature re-encoding mechanism, and the other model uses an autoencoder-based data reduction technique. The incorporation of feature re-encoding layers facilitates the QCNN model in reaching the optimal solution space more effectively, resulting in better classification accuracy. Comparative accuracy improvement is found to be higher in the Fashion MNIST dataset, suggesting that the proposed model is expected to effectively deal with complex and challenging datasets. Moreover, by combining the proposed PCA re-encoded QCNN with a conventional QCNN using autoencoder features and applying a joint optimization technique, the overall classification performance enhances significantly. It is observed that both the joint optimization and ensemble learning approaches exhibit superior performance compared to state-of-the-art methods for both the MNIST and Fashion MNIST datasets. Notably, the joint optimization strategy consistently outperforms the ensemble learning approach.

\section*{Author Contribution}
\textbf{S. M. Sarkar:} conceptualization, methodology, software, validation, formal analysis, data curation, investigation, writing—original draft, writing—review and editing. \textbf{S. I. Ahmed:} conceptualization, methodology, formal analysis, writing—review and editing, project administration. \textbf{J. Mahmud:} conceptualization, writing—review and editing, supervision. \textbf{S. A. Fattah:} writing—review and editing, supervision. \textbf{G. Sharma:} writing—review and editing, supervision. All authors reviewed the manuscript.

\section*{Competing Interest}
The authors declare no competing interests.

\bibliography{sn-bibliography}

\end{document}